\documentclass{osa-article}

\journal{osajournal}
\usepackage{bm}
\usepackage{adjustbox}
\usepackage{multirow}

\articletype{Research Article}
\begin{document}
\title{Optimizing electro-optic modulators for interfacing color centers in an integrated silicon carbide platform}

\author{Ruixuan Wang,\authormark{1} Jingwei Li,\authormark{1}, and Qing Li\authormark{*1}}

\address{\authormark{1}Department of Electrical and Computer Engineering, Carnegie Mellon University, Pittsburgh, PA 15213, USA}

\email{\authormark{*}qingli2@andrew.cmu.edu} %% email address is required



\begin{abstract}
Silicon carbide is a promising material platform for hosting various color centers suitable for quantum information processing. Here, we report the design and demonstration of an integrated electro-optic modulator that can directly interface the silicon vacancy centers in the 4H-silicon-carbide-on-insulator platform. Despite a relatively low electro-optic coefficient ($0.22$ pm/V), the optimized silicon carbide modulator is able to work in the 920 nm range with a propagation loss of less than $0.5$ dB/cm, featuring a 3-dB bandwidth around 500 MHz, an extinction ratio of 8-12 dB for an operating peak-to-peak voltage of 10 V, and a footprint of less than $0.1\ \text{mm}^2$. With such modulation performance, our work provides a cost-effective solution for the chip-scale implementation of the electro-optic modulation technology for interfacing color centers in the silicon carbide platform. 
\end{abstract}

\section{Introduction}
Color centers in solid-state materials are important resources in quantum information processing owing to their nanoscale size and potential for disruptive applications \cite{Jorg_NV_review_2013}. One prominent example is the nitrogen-vacancy (NV) center in diamonds, which has shown impressive progress over the past few decades \cite{Jorg_NV_magnetic_2008, Lunkin_NV_magnetic_2008, Yao_NV_magnetic_2024}. Recently, silicon carbide (SiC) emerged as another promising quantum material for hosting various color centers, including silicon vacancy (V$_\text{Si}$), divacancy, vanadium, etc \cite{SiC_colorcenter_review}. For example, V$_\text{Si}$ possesses a long quantum coherence time in bulk wafers, which is measured to be $>100\ \mu$s at room temperature and up to 20 ms at cryogenic temperatures \cite{SiC_colorcenter_review, Jorg_VSi_coherence_2015}. In addition, its V2 line near the wavelength of 917 nm is an encouraging candidate for ultrasensitive magnetic field and temperature sensing \cite{SiC_magnetomery_PRX, SiC_Magnetometry_2021}.  

The scalable implementation of color-center-based quantum nodes, however, necessitates a combination of chip-scale quantum technologies in a compact form \cite{Vuckovic_SiC_review}. From this perspective, SiC is advantageous compared to diamond because of its favorable material properties, including its lower cost and compatibility with complementary metal-oxide-semiconductor (CMOS) fabrication processes. Moreover, a low-loss SiC-on-insulator (SiCOI) platform has already been demonstrated, which could bring together a group of integrated functionalities on the same device platform \cite{Ou_4HSiC,Noda_4HSiC_PhC,Vuckovic_4HSiC_nphoton, Ou_4HSiC_combQ, Vuckovic_4HSiC_soliton, Li_SiC_entangled}. In this work, we will explore the addition and optimization of the electro-optic (EO) modulation technology suitable for color centers in an integrated SiC platform, which is a key building block for applications ranging from quantum sensing to communications. For instance, a high-performance EO modulator allows for encoding classical information on the quantum bits of light, or it can be employed for the pulsed shaping of the optical pump or the emitted single photons \cite{Loncar_EOM_pulse_shaping}.

So far, only SiC EO modulators focusing on classical communication in the 1550 nm band have been demonstrated \cite{Loncar_SiC_modulator, Li_SiC_EOM}, which cannot be directly applied to the quantum regime for the following reasons. First, the 1550 nm band is spectrally mismatched to the photoluminescence window of the majority of color centers found in SiC. In addition, the relatively small EO constant of SiC (on the order of 1 pm/V) often leads to a small extinction ratio in intensity modulation, below the typical minimum ER of $3.5$-8 dB required in practice \cite{Loncar_SiC_modulator, Li_SiC_EOM}. Finally, while attaining a large modulation bandwidth (BW), on the order of tens of gigahertz (GHz) \cite{Loncar_LN_MZI_modulator_Nature}, is preferred for classical communication, the long quantum coherence time of color centers renders this design goal less important. Instead, a careful balance between the available bandwidth and other performance metrics such as the extinction ratio would likely result in better overall performance.

In this work, we use the V$_\text{Si}$ defect in 4H-SiC as an example to illustrate the optimization process of electro-optic modulators (EOMs). Despite a relatively small EO coefficient possessed by 4H-SiC ($< 1$ pm/V) \cite{Li_SiC_EOM}, the optimized EOM is shown to attain a high extinction ratio of 8-12 dB while supporting GHz-level modulation bandwidth with a peak-to-peak voltage of 10 V. With a footprint of less than $0.1\ \text{mm}^2$, such compact and high-performance EOMs can directly interface color centers in an integrated SiC platform, thus lending strong support to the chip-scale implementation of quantum technologies in a cost-effective manner. 

\section{Design and optimization}

\begin{figure}[ht]
\centering
\includegraphics[width=0.8\linewidth]{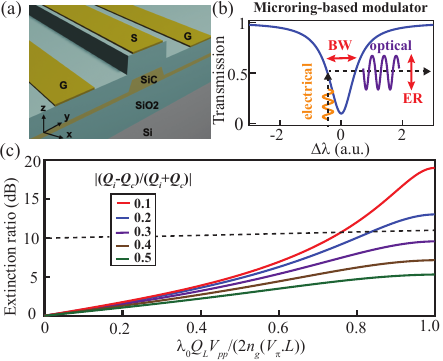}
\caption{(a) Schematic of an integrated electro-optic modulator based on a silicon carbide (SiC) waveguide (S and G denote the signal and group electrodes, respectively); (b) Illustration of the Pockels effect in a microring resonator, where the applied electrical signal modulates the optical transmission of a laser input with fixed bias. Two important performance metrics are the modulation bandwidth (BW) and the extinction ratio (ER); and (c) Expected extinction ratio as a function of the coupling condition and applied electrical voltage ($V_{pp}$ is the peak-to-peak voltage).}
\label{Fig1}
\end{figure}

As illustrated in Fig.~1a, the EOM is fabricated by placing Ti/Au electrodes on top of a SiC waveguide in the 4H-SiCOI platform, similar to our previous experiment that confirmed the Pockels effect of 4H-SiC in the 1550 nm band \cite{Li_SiC_EOM}. The wavelength range, however, is shifted to the 920 nm band to target the zero-phonon line (ZPL) of the V2 emission from V$_\text{Si}$ \cite{SiC_colorcenter_review}. In addition, given the relatively long quantum coherence time of V$_\text{Si}$ in thin-film SiC devices (even with the Purcell enhancement by coupling to nanophotonic cavities) \cite{Jorg_VSi_coherence_2015, Hu_SiC_Purcell2021}, a modulation bandwidth at the GHz level is deemed sufficient. This allows us to employ a microring resonator with a relatively high optical quality factor ($Q$), whose optical transmission is modulated by applying an electrical signal for a fixed pump laser (see Fig.~1b). 

The optimal $Q$ factor of the microresonator is determined by the trade-off between the modulation bandwidth (BW) and the achievable extinction ratio (ER). For example, the total wavelength shift ($\Delta\lambda_c$) induced by a sinusoidal signal with a peak-to-peak voltage of $V_{pp}$ can be estimated using the definition of half-wave voltage $V_\pi$, which is the voltage corresponding to a shift of half of the free spectral range (FSR). Since the FSR (in wavelength) of a microring at the resonance of $\lambda_0$ is given by $FSR=\lambda_0^2/(n_gL)$, with $n_g$ and $L$ denoting the group index and the circumference of the resonator, respectively, we obtain  
\begin{equation}
\Delta\lambda_c=\frac{V_{pp}}{V_\pi}\frac{FSR}{2}=\frac{V_{pp}}{V_\pi}\frac{\lambda_0^2}{2n_gL}.
\label{Eq_shift}
\end{equation}

In general, the achievable ER is a function of $\Delta\lambda_c$ and the coupling condition of the microresonator. For small $\Delta\lambda_c$, the maximum modulation contrast is obtained by biasing the pump laser near the half-maximum point of the transmission (see Fig.~1b). Moreover, the resonator should ideally operate near the critical coupling regime, for which the coupling $Q$ ($Q_c$) is equal to the intrinsic $Q$ ($Q_i$), resulting in an infinitely large contrast between the on- and off-resonance transmission. To quantitatively account for these two factors, we define a normalized wavelength shift as $x\equiv \Delta\lambda_c/\Delta\lambda_{FWHM}$, where $\Delta\lambda_{FWHM}$ is the full width at half maximum of the resonance ($\Delta\lambda_{FWHM}=\lambda_0/Q_L$ with $Q_L$ being the loaded $Q$ factor). As a result,
\begin{equation}
x\equiv \frac{\Delta\lambda_c}{\Delta\lambda_{FWHM}}=\frac{\lambda_0 Q_L V_{pp}}{2n_g(V_\pi\cdot L)}.
\label{Eq_x}
\end{equation}
In the above equation, the half-wave voltage $V_\pi$ and the device length $L$ is combined to form the typically quoted figure of merit of EOM, i.e., $V_\pi\cdot L$. The simulated ER corresponding to different $x$ values and coupling conditions are plotted in Fig.~1c, revealing that large ERs are only achievable when $x$ is close to 1 while simultaneously engineering $Q_c$ to be close enough to $Q_i$ (i.e., near critical coupling).

Unlike EOMs based on the Mach-Zehnder interferometer configuration, the modulation bandwidth of microresonator-based EOMs is typically limited by the photon lifetime of the cavity instead of the intrinsic RC rise/fall time \cite{Loncar_LN_ring_modulator2018}. As such, Eq.~\ref{Eq_x} also suggests a design trade-off between the ER (which is a function of $x$) and the optical modulation bandwidth (which is proportional to $\Delta\lambda_{FWHM}$) for a microresonator with a fixed $Q_i/Q_c$ ratio. In fact, we can formulate a bandwidth-extinction ratio product (BEP) from Eq.~\ref{Eq_x} as 
\begin{equation}
BEP\equiv BW\cdot x = \frac{cV_{pp}}{2n_g(V_\pi\cdot L)},
\label{Eq_BEP}
\end{equation}
where $BW$ is defined in the frequency domain as $BW=c/(\lambda_0Q_L)$ ($c$ is the speed of light in vacuum) and $x$ is the dimensionless factor defined in Eq.~\ref{Eq_x} characterizing the normalized wavelength shift. In Table 1, the performance metrics of some representative microring-based EOMs with low ERs are listed. As can be seen, those experiments prioritized bandwidth performance by employing a relatively low-$Q$ microring, which resulted in a small normalized wavelength shift (i.e., $x<<1$) and hence a suppressed ER. 

\begin{table*}[ht]
\centering
\begin{adjustbox}{width=1.2\columnwidth,center}
\begin{tabular}{c c c c c c c c c c c c}
\hline
\multirow{2}{*}{\textbf{Reference}}&\multirow{2}{*}{\textbf{Material}}&\textbf{EO coeff.}&$\bm{\lambda}_0$&\multirow{2}{*}{$\bm{n_g}$}&$\bm{Q_L}$ &$\bm{V_\pi\cdot L}$& $\bm{V_{pp}}$ & \textbf{BEP}& \textbf{BW}&\multirow{2}{*}{$\bm{x}$} &\textbf{ER}\\
& & (pm/V) & (nm) & & (k)& (V$\cdot$cm) & (V)& (GHz) & (GHz) & & (dB)\\
\hline
Wang et.al. \cite{Loncar_LN_ring_modulator2018}& LN & 30 &1550& $2.2$&8& $7.8$ & $5.66$ & $5.0$ & 30 & $0.2$ &3 \\
Powell et.al. \cite{Loncar_SiC_modulator}&3C-SiC &$1.5$& 1550 & $2.6$ & $89.3$ & 55 & 8 &$0.84$ & $2.5$ & $0.37$ & 3\\ 
\textbf{This work} & 4H-SiC& $0.22$&\textbf{920}& $2.7$& 720 & 110 & 10 & $\bm{0.5}$ & $\bm{0.45}$ & \textbf{1-1.1}& \textbf{8-12}\\   
\hline
\end{tabular}
\end{adjustbox}
\caption{Comparison of electro-optic modulators based on the Pockels effect in a microring resonator. The higher extinction ratio (ER) reported in this work for 4H-SiC is mainly achieved by optimizing the bandwidth-ER product (BEP).}
\label{Table1}
\end{table*}

\begin{figure}[ht]
\centering
\includegraphics[width=0.8\linewidth]{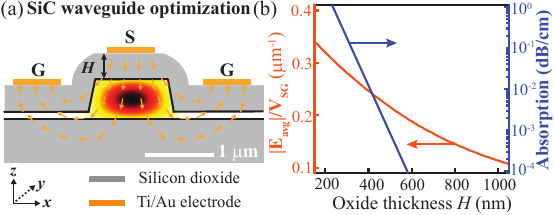}
\caption{(a) Cross-sectional view of a 4H-SiC rib waveguide with the signal (S) and ground (G) electrodes implemented near the waveguide mode (red colored). The 4H-SiC waveguide is surrounded by silicon dioxide; and (b) Simulated average z-component of the electric field per volt (red) and propagation loss (blue) due to metal absorption as a function of the oxide thickness $H$ at the wavelength of 920 nm. The width and height of the SiC waveguide are assumed to be $1.5\ \mu$m and $0.6\ \mu$m, respectively.}
\label{Fig_EO}
\end{figure}

For the above reasons, our design efforts to improve ER are focused on pushing the normalized wavelength shift $x$ towards 1 while minimizing $V_\pi\cdot L$ to boost the BEP as expressed in Eq.~\ref{Eq_BEP}. This is achieved by optimizing the geometry of a 4H-SiCOI rib waveguide with Ti/Au electrodes placed on top (see Fig.~2a). To utilize the largest Pockels coefficient of 4H-SiC, which is $r_{13}$ \cite{Li_SiC_EOM}, we employ the fundamental transverse-electric (TE$_{00}$) mode (i.e. dominant electric field in the $x-y$ plane) with the applied electric field parallel to the $c$-axis (along $z$). The induced refractive index change is given by \cite{Li_SiC_EOM}
\begin{equation}
\Delta n=-\frac{1}{2}n_o^3r_{13}\tilde{\bm E}_z, \label{Eq_dn}
\end{equation}
where $n_o$ and $r_{13}$ denote the refractive index and electro-optic coefficient of SiC, respectively, and $\tilde{\bm E}_z$ represents the $z$-component of the electric field. The corresponding resonance wavelength shift can be derived using the following relationship:
\begin{align}
&\Delta \lambda_c=\frac{\lambda_0 n_o^2\Gamma}{2}|r_{13}<\tilde{\bm E}_z,V_{pp}>|, \label{Eq_r13}
\end{align}
where $\Gamma$ is the mode confinement factor and $<\tilde{\bm E}_z,V_{pp}>|$ is the averaged electric field in the $z$-direction corresponding to $V_{pp}$.

For efficient EO tuning, the average $z$ component of the electric field induced by a unit voltage is calculated as a function of the thickness of the top oxide cladding layer $H$ (see Fig.~2b). Meanwhile, one needs to be aware of the optical absorption from the metal pads if $H$ is too small. The simulation results shown in Fig.~\ref{Fig_EO}b point to the optimal thickness of the oxide cladding in the range of $400-450$ nm, for which the estimated absorption loss ($<0.01$ dB/cm) is considered low enough at 920 nm. With this reduced oxide thickness, the averaged electrical field is increased by a factor of $2.5$-3 compared to our previous design that employed 1-$\mu$m-thick oxide for the EOM demonstration in the 1550 nm band \cite{Li_SiC_EOM}. This improvement is expected to proportionally reduce the required $V_{pp}$ for the EOM operation.

\section{Experimental demonstration}

\begin{figure}[ht]
\centering
\includegraphics[width=0.8\linewidth]{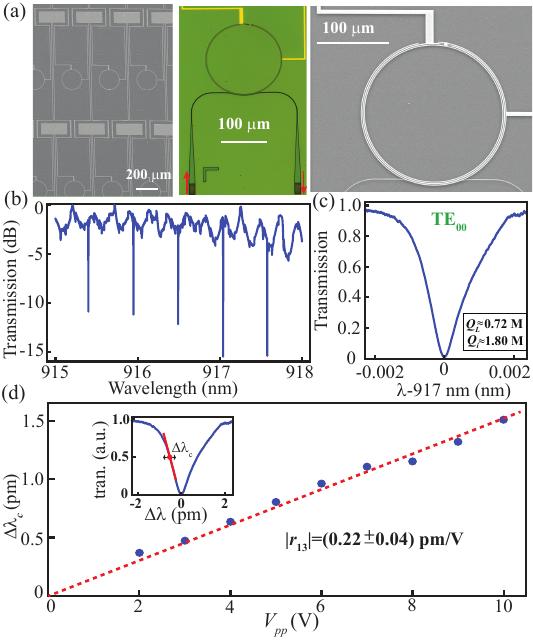}
\caption{(a) Left: Scanning electro micrograph (SEM) of the fabricated SiC microring array with electrical pads; Middle: Optical micrograph of a 90-$\mu$m-radius SiC microring with grating couplers; Right: SEM image of the same SiC microring shown in the middle panel; (b) Measured swept-wavelength transmission of the SiC microring in (a) in the 920 nm band; (c) Zoom-in figure of the select transverse-electric fundamental resonance (TE$_{00}$) near 917 nm, displaying a loaded ($Q_L$) and intrinsic ($Q_i$) quality factors of $0.72$ million and $1.80$ million, respectively. and (d) Measured resonance wavelength shifts (blue dots) and the corresponding linear fit (red solid line) for varied peak-to-peak voltages ($V_{pp}$) of a sinusoidal electrical signal at a fixed frequency of 10 MHz.}
\label{Fig3}
\end{figure}
Our pattern for the experimental demonstration consists of an array of EOMs with varied parameters to simultaneously meet the requirements of a large normalized wavelength shift ($x\approx 1$), an optimized $V_\pi \cdot L$, and a critical coupling condition for the select resonant mode. As shown in Fig.~3a, each EOM includes a ring resonator with a radius of 90 $\mu$m, a pair of grating couplers that serve as optical input and output ports, and Ti/Au electrodes to apply electrical signals. In terms of critical coupling engineering, the ring waveguide width is fixed at $1.5\ \mu$m, while the width of the bus waveguide is swept around a nominal value of $0.88\ \mu$m to achieve phase matching to the TE$_{00}$ mode for a fixed gap of 150 nm and a pulley coupling length of $70\ \mu$m (see the scanning electron micrograph shown in Fig.~3a) \cite{Li_FWMBS}. 

The nanofabrication begins with a semi-insulating 600-nm-thick 4H-SiCOI chip obtained from a customized bonding and polishing process (NGK Insulators). The pattern is defined using electron beam lithography (EBL) with 1-$\mu$m-thick flowable oxide (FOx) resist, which is subsequently transferred to the SiC layer through dry etching \cite{Li_4HSiC_comb}. The etch depth is controlled to be approximately $500$ nm, leaving an unetched SiC layer (pedestal) of 100 nm. Next, we deposit a layer of $450$-nm-thick silicon oxide cladding on top of SiC devices via plasma-enhanced chemical vapor deposition (PECVD). Following that, electrodes made of Ti/Au layers ($30 $ nm Ti and $70 $ nm Au) are added using a lift-off process consisting of EBL and electron-beam evaporation. As shown in Fig.~3a, the electrodes feature a tight spacing of 1 $\mu$m to the waveguide sidewalls to maximize the overlap between the optical and electrical fields. 

\begin{figure}[ht]
\centering
\includegraphics[width=0.7\linewidth]{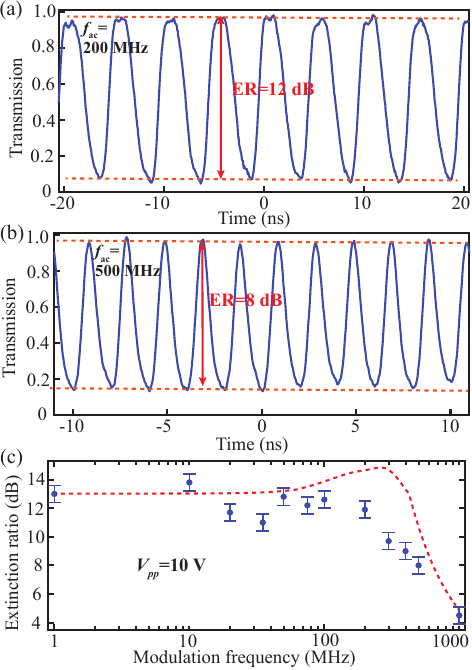}
\caption{(a) Recorded optical transmission of the TE$_{00}$ resonance shown in Fig.~3(c) with a modulation frequency of 200 MHz for a fixed $V_{pp}=10$ V; (b) Similar to (a) but with an increased modulation frequency of 500 MHz; and (c) Summarized extinction ratio (blue dots) as a function of the modulation frequency for a fixed $V_{pp}=10$ V. The error bars are to describe the measurement uncertainties, while the red dashed line is the simulation result.}
\label{Fig4}
\end{figure}

The experimental setup used in the EO modulator characterization supports the optical measurements of silicon carbide microrings through grating couplers, as well as electro-optic tuning through a high-performance microwave (GGB) probe.  For example, the optical linear transmission measurement is performed using a narrow-linewidth laser (Msquared, linewidth $<100$ kHz), which has a tunable wavelength range of 700-1000 nm. In addition, its TeraScan feature enables automated wavelength scans within a terahertz range. The grating coupler is measured to show a coupling loss around 7-9 dB \cite{Li_4HSiC_comb}. The linear transmission displayed in Fig.~3b indicates that our pulley coupling scheme has excited only one dominant mode family, which is confirmed to be TE$_{00}$ by comparing the measured FSRs (around 200 GHz) to simulation. Finally, the close-up view of the 917 nm resonance, shown in Fig.~3c, reveals an intrinsic $Q$ of $1.8$ million and a loaded $Q$ near $0.72$ million, corresponding to an on-off resonance contrast of more than 14 dB. We note that this level of intrinsic $Q$ ($1.8$ million), equivalent to an intrinsic propagation loss of $0.44$ dB/cm, is the highest reported for 4H-SiC devices in the 920 nm range \cite{Vuckovic_4HSiC_nphoton}.

To quantify the EO response, a $10$-MHz sinusoidal electrical signal is applied to the EOM through the microwave probe while biasing the pump laser at the midpoint of the resonance transmission. The resulting resonance wavelength shift is inferred from the optical transmission and plotted in Fig.~2d. With the help of Eq.~\ref{Eq_r13}, we can extract the magnitude of the EO coefficient $r_{13}$ to be ($0.22\pm 0.04$) pm/V, consistent with our previous measurements performed on Cree 4H-SiC wafers \cite{Li_SiC_EOM}. In addition, the slope of $\Delta\lambda_c$ versus the applied $V_{pp}$ can be used to estimate $V_\pi \cdot L$ based on Eq.~\ref{Eq_shift}. As listed in Table 1,  a decent $V_\pi \cdot L$ around $110$ V$\cdot$cm is obtained for our 4H-SiC EOM, which is only 2x of that obtained in 3C-SiC despite a significantly lower EO coefficient (7x difference) \cite{Loncar_SiC_modulator}. This relative strength is mostly attributed to the optimized PECVD oxide thickness and precise alignment of the electrodes with respect to the optical mode. By employing a different type of 4H-SiC (e.g., the $r_{13}$ coefficient of Norstel 4H-SiC is at least twice that of the Cree wafer), it is feasible to further reduce $V_\pi \cdot L$ below $50$ V$\cdot$ cm \cite{Li_SiC_EOM}.

The performance of our EOM is further tested at higher modulation frequencies using a high-bandwidth photodetector (Newport AD-40ir with 12 GHz bandwidth) and a high-speed oscilloscope (Keysight DSO-X 6002A with 1 GHz bandwidth). For example, Figs.~4a and 4b showcase two examples of optical signals recorded for a fixed $V_{pp}$ of 10 V. As can be seen, for the modulation frequency of 200 MHz (Fig.~4a), the ER is measured to be approximately 12 dB, for which the normalized wavelength shift $x$ is estimated near $1.1$. When increasing the modulation frequency to 500 MHz (Fig.~4b), which is slightly above the cavity linewidth (450 MHz), the ER drops to 8 dB as expected. In Fig.~4c, the experimentally measured ERs at various modulation frequencies are compared against the simulation, for which reasonable agreement has been achieved. The demonstrated ER levels, which are above 8 dB for modulation frequencies up to 500 MHz, could enable a host of applications including pulsed shaping of the optical pump or the emitted single photons, as well as encoding the classical information on the quantum bits through intensity modulation \cite{Loncar_EOM_pulse_shaping}. 
\clearpage
\section{Conclusion}
In conclusion, we have designed and demonstrated a high-performance electro-optic intensity modulator that could interface silicon vacancy defects directly in an integrated 4H-SiCOI platform. The modulator operates in the 920 nm wavelength range, and features a modulation bandwidth of 450 MHz, an extinction ratio of 8-12 dB, and a footprint of $<0.1 \text{mm}^2$. These performance metrics are achieved by optimizing the EOM geometry and the modal overlap between the electrical and optical fields, despite a relatively weak electro-optic coefficient possessed by 4H-SiC. As such, we believe our work paves the way for the chip-scale implementation of electro-optic modulation technologies in the 4H-SiC platform without requiring complicated heterogeneous integration. In addition, our approach can be extended to other wavelengths that are compatible with different types of color centers discovered in SiC. 
\begin{backmatter}
\bmsection{Funding}
This work was supported by NSF (2240420). 

\bmsection{Acknowledgments}
The authors acknowledge the use of Bertucci Nanotechnology Laboratory at Carnegie Mellon University supported by grant BNL-78657879 and the Materials Characterization Facility supported by grant MCF-677785. R.~Wang and J.~Li also acknowledge the support of Tan Endowed Graduate Fellowship and Benjamin Garver Lamme/Westinghouse Graduate Fellowship from CMU, respectively. 

\bmsection{Disclosures}  The authors declare no conflicts of interest.

\bmsection{Data Availability} Data underlying the results presented in this paper are not publicly available at this time but may be obtained from the authors upon reasonable request.

\end{backmatter}
\bibliography{SiC_Ref}

\begin{thebibliography}{10}
\newcommand{\enquote}[1]{``#1''}

\bibitem{Jorg_NV_review_2013}
M.~W. Doherty, N.~B. Manson, P.~Delaney, F.~Jelezko, J.~Wrachtrup, and L.~C. Hollenberg, \enquote{The nitrogen-vacancy colour centre in diamond,} {\protect\JournalTitle{Physics Reports}} \textbf{528}, 1--45 (2013).

\bibitem{Jorg_NV_magnetic_2008}
G.~Balasubramanian, I.~Y. Chan, R.~Kolesov, M.~Al-Hmoud, J.~Tisler, C.~Shin, C.~Kim, A.~Wojcik, P.~R. Hemmer, A.~Krueger, T.~Hanke, A.~Leitenstorfer, R.~Bratschitsch, F.~Jelezko, and J.~Wrachtrup, \enquote{Nanoscale imaging magnetometry with diamond spins under ambient conditions,} {\protect\JournalTitle{Nature}} \textbf{455}, 648--651 (2008).

\bibitem{Lunkin_NV_magnetic_2008}
J.~R. Maze, P.~L. Stanwix, J.~S. Hodges, S.~Hong, J.~M. Taylor, P.~Cappellaro, L.~Jiang, M.~V.~G. Dutt, E.~Togan, A.~S. Zibrov, A.~Yacoby, R.~L. Walsworth, and M.~D. Lukin, \enquote{Nanoscale magnetic sensing with an individual electronic spin in diamond,} {\protect\JournalTitle{Nature}} \textbf{455}, 644--647 (2008).

\bibitem{Yao_NV_magnetic_2024}
P.~Bhattacharyya, W.~Chen, X.~Huang, S.~Chatterjee, B.~Huang, B.~Kobrin, Y.~Lyu, T.~J. Smart, M.~Block, E.~Wang, Z.~Wang, W.~Wu, S.~Hsieh, H.~Ma, S.~Mandyam, B.~Chen, E.~Davis, Z.~M. Geballe, C.~Zu, V.~Struzhkin, R.~Jeanloz, J.~E. Moore, T.~Cui, G.~Galli, B.~I. Halperin, C.~R. Laumann, and N.~Y. Yao, \enquote{Imaging the {Meissner} effect in hydride superconductors using quantum sensors,} {\protect\JournalTitle{Nature}} \textbf{627}, 73--79 (2024).

\bibitem{SiC_colorcenter_review}
S.~Castelletto and A.~Boretti, \enquote{Silicon carbide color centers for quantum applications,} {\protect\JournalTitle{Journal of Physics: Photonics}} \textbf{2}, 022001 (2020).

\bibitem{Jorg_VSi_coherence_2015}
M.~Widmann, S.-Y. Lee, T.~Rendler, N.~T. Son, H.~Fedder, S.~Paik, L.-P. Yang, N.~Zhao, S.~Yang, I.~Booker, A.~Denisenko, M.~Jamali, S.~A. Momenzadeh, I.~Gerhardt, T.~Ohshima, A.~Gali, E.~Janzén, and J.~Wrachtrup, \enquote{Coherent control of single spins in silicon carbide at room temperature,} {\protect\JournalTitle{Nature Materials}} \textbf{14}, 164--168 (2015).

\bibitem{SiC_magnetomery_PRX}
D.~Simin, V.~Soltamov, A.~Poshakinskiy, A.~Anisimov, R.~Babunts, D.~Tolmachev, E.~Mokhov, M.~Trupke, S.~Tarasenko, A.~Sperlich, P.~Baranov, V.~Dyakonov, and G.~Astakhov, \enquote{All-{Optical} dc {Nanotesla} {Magnetometry} {Using} {Silicon} {Vacancy} {Fine} {Structure} in {Isotopically} {Purified} {Silicon} {Carbide},} {\protect\JournalTitle{Physical Review X}} \textbf{6}, 031014 (2016).

\bibitem{SiC_Magnetometry_2021}
J.~B.~S. Abraham, C.~Gutgsell, D.~Todorovski, S.~Sperling, J.~E. Epstein, B.~S. Tien-Street, T.~M. Sweeney, J.~J. Wathen, E.~A. Pogue, P.~G. Brereton, T.~M. McQueen, W.~Frey, B.~D. Clader, and R.~Osiander, \enquote{Nanotesla {Magnetometry} with the {Silicon} {Vacancy} in {Silicon} {Carbide},} {\protect\JournalTitle{Physical Review Applied}} \textbf{15}, 064022 (2021).

\bibitem{Vuckovic_SiC_review}
D.~M. Lukin, M.~A. Guidry, and J.~Vu{\v c}kovi{\'c}, \enquote{Integrated {{quantum photonics}} with {{silicon carbide}}: challenges and {{prospects}},} {\protect\JournalTitle{PRX Quantum}} \textbf{1}, 020102 (2020).

\bibitem{Ou_4HSiC}
Y.~Zheng, M.~Pu, A.~Yi, X.~Ou, and H.~Ou, \enquote{{{4H}}-{{SiC}} microring resonators for nonlinear integrated photonics,} {\protect\JournalTitle{Optics Letters}} \textbf{44}, 5784 (2019).

\bibitem{Noda_4HSiC_PhC}
B.-S. Song, T.~Asano, S.~Jeon, H.~Kim, C.~Chen, D.~D. Kang, and S.~Noda, \enquote{Ultrahigh-{{Q}} photonic crystal nanocavities based on {{4H}} silicon carbide,} {\protect\JournalTitle{Optica}} \textbf{6}, 991 (2019).

\bibitem{Vuckovic_4HSiC_nphoton}
D.~M. Lukin, C.~Dory, M.~A. Guidry, K.~Y. Yang, S.~D. Mishra, R.~Trivedi, M.~Radulaski, S.~Sun, D.~Vercruysse, G.~H. Ahn, and J.~Vu{\v c}kovi{\'c}, \enquote{{{4H}}-silicon-carbide-on-insulator for integrated quantum and nonlinear photonics,} {\protect\JournalTitle{Nature Photonics}} \textbf{14}, 330--334 (2020).

\bibitem{Ou_4HSiC_combQ}
C.~Wang, Z.~Fang, A.~Yi, B.~Yang, Z.~Wang, L.~Zhou, C.~Shen, Y.~Zhu, Y.~Zhou, R.~Bao, Z.~Li, Y.~Chen, K.~Huang, J.~Zhang, Y.~Cheng, and X.~Ou, \enquote{High-{{Q}} microresonators on {{4H}}-silicon-carbide-on-insulator platform for nonlinear photonics,} {\protect\JournalTitle{Light: Science \& Applications}} \textbf{10}, 139 (2021).

\bibitem{Vuckovic_4HSiC_soliton}
M.~A. Guidry, D.~M. Lukin, K.~Y. Yang, R.~Trivedi, and J.~Vučković, \enquote{Quantum optics of soliton microcombs,} {\protect\JournalTitle{Nature Photonics}} \textbf{16}, 52--58 (2022).

\bibitem{Li_SiC_entangled}
A.~Rahmouni, L.~Ma, R.~Wang, J.~Li, X.~Tang, T.~Gerrits, Q.~Li, and O.~Slattery, \enquote{Entangled photon pair generation in an integrated silicon carbide platform,} {\protect\JournalTitle{Research Square}} p. 3069754 (2023).

\bibitem{Loncar_EOM_pulse_shaping}
D.~Zhu, C.~Chen, M.~Yu, L.~Shao, Y.~Hu, C.~J. Xin, M.~Yeh, S.~Ghosh, L.~He, C.~Reimer, N.~Sinclair, F.~N.~C. Wong, M.~Zhang, and M.~Lončar, \enquote{Spectral control of nonclassical light pulses using an integrated thin-film lithium niobate modulator,} {\protect\JournalTitle{Light: Science \& Applications}} \textbf{11}, 327 (2022).

\bibitem{Loncar_SiC_modulator}
K.~Powell, L.~Li, A.~Shams-Ansari, J.~Wang, D.~Meng, N.~Sinclair, J.~Deng, M.~Lončar, and X.~Yi, \enquote{Integrated silicon carbide electro-optic modulator,} {\protect\JournalTitle{Nature Communications}} \textbf{13}, 1851 (2022).

\bibitem{Li_SiC_EOM}
R.~Wang, J.~Li, L.~Cai, and Q.~Li, \enquote{Investigation of the electro-optic effect in high-{{Q}} {{4H-SiC}} microresonators,} {\protect\JournalTitle{Optics Letters}} \textbf{48}, 1482--1485 (2023).

\bibitem{Loncar_LN_MZI_modulator_Nature}
C.~Wang, M.~Zhang, X.~Chen, M.~Bertrand, A.~Shams-Ansari, S.~Chandrasekhar, P.~Winzer, and M.~Lončar, \enquote{Integrated lithium niobate electro-optic modulators operating at {CMOS}-compatible voltages,} {\protect\JournalTitle{Nature}} \textbf{562}, 101--104 (2018).

\bibitem{Hu_SiC_Purcell2021}
M.~N. Gadalla, A.~S. Greenspon, R.~K. Defo, X.~Zhang, and E.~L. Hu, \enquote{Enhanced cavity coupling to silicon vacancies in {4H} silicon carbide using laser irradiation and thermal annealing,} {\protect\JournalTitle{Proceedings of the National Academy of Sciences}} \textbf{118} (2021).

\bibitem{Loncar_LN_ring_modulator2018}
C.~Wang, M.~Zhang, B.~Stern, M.~Lipson, and M.~Lončar, \enquote{Nanophotonic lithium niobate electro-optic modulators,} {\protect\JournalTitle{Optics Express}} \textbf{26}, 1547--1555 (2018).

\bibitem{Li_FWMBS}
Q.~Li, M.~Davan{\c c}o, and K.~Srinivasan, \enquote{Efficient and low-noise single-photon-level frequency conversion interfaces using silicon nanophotonics,} {\protect\JournalTitle{Nature Photonics}} \textbf{10}, 406--414 (2016).

\bibitem{Li_4HSiC_comb}
L.~Cai, J.~Li, R.~Wang, and Q.~Li, \enquote{Octave-spanning microcomb generation in {4H}-silicon-carbide-on-insulator photonics platform,} {\protect\JournalTitle{Photonics Research}} \textbf{10}, 870--876 (2022).

\end{thebibliography}
\end{document}